\documentclass[twocolumn]{revtex4}

\usepackage{amsmath}
\usepackage{amsfonts}
\usepackage{amssymb}
\usepackage{graphicx}
\usepackage{rotating}
\allowdisplaybreaks
\usepackage{setspace}
\usepackage{bm}
\usepackage{color}

\begin{document}

\title{Excitation Variance Matching with Limited Configuration Interaction Expansions in Variational Monte Carlo}

\author{Paul J. Robinson,${^1}$ Sergio D. Pineda Flores,${^2}$ and Eric Neuscamman$^{2,3,}$\footnote[1]{Electronic mail: eneuscamman@berkeley.edu}}
\affiliation{$^1$Department of Chemistry and Biochemistry, University of California, Los Angeles, California 90095, USA\\
             $^2$Department of Chemistry, University of California, Berkeley, California 94720, USA\\
             $^3$Chemical Sciences Division, Lawrence Berkeley National Laboratory, Berkeley, California 94720, USA}

\begin{abstract}
In the regime where traditional approaches to electronic structure cannot afford to
achieve accurate energy differences via exhaustive wave function flexibility,
rigorous approaches to balancing different states' accuracies become desirable.
As a direct measure of a wave function's accuracy, the energy variance offers one route
to achieving such a balance.
Here, we develop and test a variance matching approach for predicting excitation energies
within the context of variational Monte Carlo and selective configuration interaction.
In a series of tests on small but difficult molecules, we demonstrate that the approach
it is effective at delivering accurate excitation energies when the wave function
is far from the exhaustive flexibility limit.
Results in C$_3$,
where we combine this approach with variational Monte Carlo orbital optimization,
are especially encouraging.
\end{abstract}

\maketitle


\section{Introduction}
\label{sec:intro}

Configuration interaction (CI) \cite{MolElecStruc}, in which a many-electron wave function is approximated as a finite linear
combination of Slater determinants, is among the most venerable and widely used approaches for predicting the effects of
electron correlation on molecules' properties and behavior.
The use and development of CI-based theories continues today
\cite{ Sherrill:2005:importConfigs, Ruedenberg:2009:deadwood, booth2011CarbonDimer, werner2011icmrci, blunt2015krylov, blunt2015semi, Knowles:2015:compressive, Alavi:2016:fciqmc_casscf,
       Hoffmann:2016:ici, Evangelista2016GuaranteedAcc, Tubman:2016:asci, Holmes:2016:heatBath, sharma2017semistochastic, zimmerman2017incremental, ohtsuka2017selected}
despite their central shortcoming of having to choose between being size-inconsistent
(in truncated forms) or having a cost that scales factorially with system size (in full (FCI) or complete active space (CAS) forms).
Indeed, many researchers are willing to live with this flaw in light of the stability and robustness that arise from CI's
variational nature and straightforward systematic improvability.

Recently, these qualities have led to a resurgence in the study of selective CI (sCI), in which only a parsimonious selection of
determinants is chosen, rather than the more traditional selection of all determinants below a given excitation level.
sCI methods were originally developed
\cite{bender1969studies,huron1973iterative,buenker1974individualized,gouyet1976spin,Evangelisti:1983hf,harrison1991approximating}
at a time when high-accuracy treatments of weak correlation were being sought,
a role that has now been more or less filled by coupled cluster theory. \cite{Barlett:2007:cc_rev}
More recently, there has been much renewed interest in sCI
\cite{Sherrill:2005:importConfigs, Ruedenberg:2009:deadwood, Knowles:2015:compressive, Hoffmann:2016:ici, Evangelista2016GuaranteedAcc,
      Tubman:2016:asci, Holmes:2016:heatBath, sharma2017semistochastic, zimmerman2017incremental, ohtsuka2017selected},
especially in the context of treating strong correlation.
While these approaches have made impressive progress, they do not remove the fundamental challenge that any CI,
selective or otherwise, will lose its ability to systematically converge as system size increases and the number of determinants
needed for a given level of accuracy quickly overwhelms available computing resources.
Applications of CI, and indeed many other wave functions, to the wide class of chemical systems in which dozens or hundreds of
atoms are involved thus require methods that can succeed without resorting to an exhaustive expansion of wave function flexibility.

In the context of predicting energy differences, one might instead rely on a cancellation of errors.
In practice, however, this approach can be frustrated by the difficulty of achieving a balance
between different states' accuracies.
At present, the most effective methods for achieving the necessary balance are those based on the state-averaged complete active space
self-consistent field (CASSCF) approach, \cite{Werner:1985:mcscf,Finley:1998:multistate_caspt2} but these rely on the exhaustive
inclusion of all active space configurations and often the singles and doubles excitations out of the active space as well.
Even difference dedicated CI, 
\cite{garcia1995iterative,garcia1998treatment,neese2003spectroscopy,Zimmerman:2017:ddci_sf}
which aims to achieve balance without resorting to a full CAS, requires a relatively exhaustive inclusion of all wave function
components related to the change between two states.
In cases where even this more limited exhaustion of flexibility is not feasible, or where the physical insight required to choose
the minimal CAS is lacking, a different route to achieving balance is required.


Here, we propose to achieve balanced treatments and accurate excitation energies
in very short sCI expansions by employing the energy variance $\sigma^2$ in combination with
recent advances in state-specific optimization within variational Monte Carlo (VMC). \cite{Zhao:2016:dir_tar}
Like Evangelista's ``guaranteed accuracy'' measure, \cite{Evangelista2016GuaranteedAcc}
$\sigma^2$ is a direct measure of wave function accuracy.
By varying different states' sCI expansion lengths so that they are of equal accuracy
as measured by $\sigma^2$, we will show that effective error cancellation and accuracies in the range
of 0.1 or 0.2 eV can be achieved even for very short sCI expansions.
Of course, this approach does not remove the ability to seek systematic convergence, as the variance-matched
energy differences are guaranteed to converge to the exact result as the sCI expansions are enlarged.
Instead, the approach allows for highly accurate results to be achieved even when one is far from the
exact $\sigma^2=0$ limit.

Certainly state-specific optimizations in VMC are not the only place where $\sigma^2$ may be useful
for achieving balance, and
one can ask whether the approach is compatible with other Monte Carlo methods,
such as full configuration interaction QMC \cite{booth2009fermion,blunt2015semi} and Monte Carlo CI.
\cite{greer1995estimating,greer1998monte}
In the former, an unbiased evaluation of $\sigma^2$ may be possible based on the same two-population system
used for density matrices. \cite{overy2014unbiased}
In the latter, it may be possible to estimate $\sigma^2$ using a similar approach as is currently used
to make perturbative corrections to sCI energies.


For the VMC approach pursued here, it is important to stress that two important advantages mitigate the fact
that achievable expansion lengths are much shorter in VMC as compared to many other sCI approaches.
First, Jastrow factors are known to significantly reduce the number of
Slater determinants required for an accurate CI expansion. \cite{UmrTouFilSorHen-PRL-07}
Second, these multi-Slater Jastrow (MSJ) wave functions (whose computational efficiency has recently been greatly enhanced
\cite{miguel:2011:table,Scuseria:2012:msj,Filippi:2016:msj_gradients,assaraf2017optimizing})
can be used to provide nodal surfaces for diffusion Monte Carlo (DMC), \cite{Foulkes:2001:qmc_review}
which is highly effective at recovering the effects of weak correlation that can be missed by limited sCI expansions.
Crucially, both MSJ wave functions \cite{assaraf2017optimizing} and DMC \cite{benali2014application}
have been demonstrated to scale to systems with over 100 atoms,
implying that the advantages of variance matching should scale to a wide variety of chemical applications.
In this paper, we will take a first step towards such applications by exploring the practical considerations that
arise in working with this idea and by establishing its accuracy in a handful of small systems where high level
benchmarks are available.


\section{Theory}
\label{sec:theory}

\subsection{Estimating the variance}
\label{sec::estimate_variance}

The energy variance, 
\begin{align}
\sigma^2 = \frac{\langle\Psi| ( \hat{H} - E )^2 |\Psi\rangle}{\langle\Psi|\Psi\rangle},
\label{eqn:var}
\end{align}
can be estimated by Monte Carlo
integration following the placement of a resolution of the identity in
between the two factors of $(\hat{H} - E)$,
\begin{align}
\sigma^2 = \frac{1}{\langle\Psi|\Psi\rangle} \int \left| \langle\vec{r}\hspace{0.5mm}| ( \hat{H} - E ) |\Psi\rangle \right|^2 d\vec{r}.
\label{eqn:var_int}
\end{align}
Defining the local energy as
$E_L(\vec{r}\hspace{0.5mm})=\langle\vec{r}\hspace{0.5mm}| \hat{H} |\Psi\rangle/\langle\vec{r}\hspace{0.5mm}|\Psi\rangle$
and letting $\Psi(\vec{r}\hspace{0.5mm})=\langle\vec{r}\hspace{0.5mm}|\Psi\rangle$,
this integral can be written as the square deviation of the local versus total energy
averaged over the wave function's probability distribution,
\begin{align}
\sigma^2 = \int \frac{\left|\Psi(\vec{r}\hspace{0.5mm})\right|^2}{\langle\Psi|\Psi\rangle} \hspace{0.7mm} \big| \hspace{0.4mm} E_L(\vec{r}\hspace{0.5mm}) - E \hspace{0.4mm} \big|^2 d\vec{r}.
\label{eqn:var_mc_int}
\end{align}
Estimating $\sigma^2$ by VMC thus amounts to sampling from the wave function's probability distribution
and averaging this square energy difference, which requires that exactly the same information be
extracted from the wave function ansatz as in a standard VMC estimation of the energy $E$. \cite{Foulkes:2001:qmc_review,Umrigar:2005:lm}
Crucially, the resolution of the identity between the two powers of $\hat{H}$ ensures that $\sigma^2$ can be estimated
without ever squaring the Hamiltonian operator explicitly.
This advantage is the reason that VMC can access the variance more easily than more traditional methods in quantum chemistry,
such as configuration interaction \cite{MolElecStruc} or coupled cluster theory. \cite{Barlett:2007:cc_rev}
Indeed, given that VMC energy estimates also require evaluating $E_L$ at a large number of sampled configurations,
an estimate of the variance is typically available ``for free'' during the normal use of VMC for energy evaluation and minimization.
However, as we discuss in Section \ref{sec::nodeless_guiding}, it can be statistically advantageous to estimate the integral
in Eq.\ (\ref{eqn:var_int}) via Monte Carlo integration over a different probability distribution than that of the wave function.

\subsection{Targeting excited states}
\label{sec::target_excited}

To optimize a MSJ expansion for a particular Hamiltonian eigenstate,
we minimize the recently introduced \cite{Zhao:2016:dir_tar} target function
\begin{align}
\Omega(\Psi) \equiv \frac{\langle\Psi|(\omega-H)|\Psi\rangle}{\langle\Psi|(\omega-H)^2|\Psi\rangle}
= \frac{\omega-E}{(\omega-E)^2 + \sigma^2}
\label{eqn:es_var_func}
\end{align}
(in which the second equality follows from $\sigma^2=\langle H^2 \rangle - E^2$)
with respect to both the nonlinear Jastrow factor variables and the CI expansion's linear coefficients.
Similar to the energy function, whose global minimum is the ground state, this function has its global minimum at the
energy eigenstate whose energy is immediately above the target energy $\omega$, allowing it to target either ground
states or excited states.
As with the variance $\sigma^2$, the denominator of $\Omega$ may be evaluated efficiently \cite{Zhao:2016:dir_tar}
by resolving an identity between the two powers of $\hat{H}$ and employing Monte Carlo integration.
As we have done in previous uses of this excited state target function, \cite{Zhao:2016:dir_tar,Neuscamman:2016:var,Zhao:2017:blocked_lm}
we ensure a unique choice for $\omega$ by adjusting it to minimize $\Omega(\Psi)$ for the
state in question.

\subsection{Balancing the states}
\label{sec::balancing}

At present, the most effective approaches for achieving a balanced description of ground and excited states
are exhaustive in nature.
For example, CAS-based approaches such as MRCI+Q \cite{werner2011icmrci} rely on the balance provided by constructing the
active space to contain all configurations that make major contributions to either state.
While these approaches are highly effective, this type of exhaustive strategy is not feasible in larger systems,
for which it is natural to ask the question:  what is the best alternative?
One could try to give ``equal'' flexibility to both states' ansatzes in a sCI by allowing them the same number
of configuration state functions (CSFs), although there is no reason to assume that this CSF-matched approach will
yield states with similar energy errors.
Instead, we would prefer to seek cancellation of error by matching a quantity that is actually related to the error
in a wave function.

In this study, we will therefore explore the possibility of improving accuracy in energy differences by insisting
that such differences be taken between wave functions with equal variance, as $\sigma^2$ is a direct measure of wave function accuracy.
We take as our wave function a MSJ expansion,
\begin{align}
\Psi(\vec{r}\hspace{0.5mm}) = e^{U(\vec{r}\hspace{0.5mm})} \sum_I^{N} \Phi_I(\vec{r}\hspace{0.5mm})
\label{eqn:msj_wfn}
\end{align}
in which each $\Phi_I$ is a different CSF --- i.e.\ a spin adapted linear combination of determinants with the same
spinless occupation pattern --- and the factor outside the sum is the Jastrow factor. \cite{Foulkes:2001:qmc_review}
In this study, the Jastrow factor is taken as
\begin{align}
U(\vec{r}\hspace{0.5mm}) = \sum_{i p} V_p(r_{ip}) + \sum_{i<j} W(r_{ij}),
\label{eqn:jastrow}
\end{align}
in which $r_{ip}$ is the distance between the $i$th electron and $p$th nucleus,
$r_{ij}$ the distance between the $i$th and $j$th electrons,
and $V_p$ and $W$ are one-dimensional functions represented by optimizable 10-point splines.
Note that we use two different functions for $W$ for same-spin versus opposite-spin electron pairs
and enforce the relevant electron-electron cusp condition in both cases.

Variational optimization of Eq.\ (\ref{eqn:msj_wfn}) via the minimization of Eq.\ (\ref{eqn:es_var_func})
will produce one value of the variance for each length $N$ chosen for the CSF expansion.
To assist in comparing the variances (or energies) of different states, we will often interpolate by fitting this
$\sigma^2$ (or $E$) versus $N$ distribution to a power law decay of the form
\begin{align}
\sigma^2(N) \approx c + \frac{d}{N^\alpha}
\label{eqn:power_law_fit}
\end{align}
in which $c$, $d$, and $\alpha$ are chosen by a least-squares fit to the $\sigma^2$ values of the
optimized wave functions for a handful of different CSF expansion lengths $N$.
This approach allows us to avoid tedious searches for the precise expansion length $N$ at which
the excited state variance most closely matches that of the ground state.

Using these interpolating functions, our approach to variance matching is the following.
First, we optimize the ground state for a small CSF expansion and evaluate its variance.
We then optimize the excited state in question for a range of expansion lengths and interpolate
via Eq.\ (\ref{eqn:power_law_fit}) to find the excited state expansion length for which the ground
and excited states' variances match.
We then employ an analogous interpolation of the excited state energy to produce the
variance-matched excitation energy.

\subsection{Modified guiding function}
\label{sec::nodeless_guiding}

Unfortunately, the statistical uncertainty of VMC estimates of the variance are not well defined
\cite{Trail:2008:heavy_tail,Trail:2008:alt_sampling} when using the standard
$|\Psi(\vec{r}\hspace{0.5mm})|^2$ importance sampling
function for the Monte Carlo integration evaluation of Eq.\ (\ref{eqn:var_int}).
To understand why this comes about, consider the nodal surface of the wave function.
Near a node, the wave function is by definition heading to a zero value, but its second derivative,
which contributes to the local energy via the kinetic energy term, need not be zero.
It is thus possible for 
\begin{align}
\left|E_L(\vec{r}\hspace{0.5mm})\right|
\hspace{1.0mm} \rightarrow \hspace{1.0mm} \left|\frac{\nabla^2\Psi}{\Psi(\vec{r}\hspace{0.5mm})}\right|
\hspace{1.0mm} \rightarrow \hspace{1.0mm} \infty
\label{eqn:local_e_divergence}
\end{align}
as $\vec{r}$ approaches the nodal surface.
While the average energy and its variance remain finite despite this divergence,
the variance of the variance
does not, and so the central limit theorem cannot be used to estimate the statistical uncertainty
in the estimated value of the variance.
Here, we will address this issue by modifying the guiding function so as
to avoid problematic divergences
(see Appendix A for an explicit example of how such modifications can help).

\begin{figure*}
\centering
\includegraphics[width=0.8\textwidth]{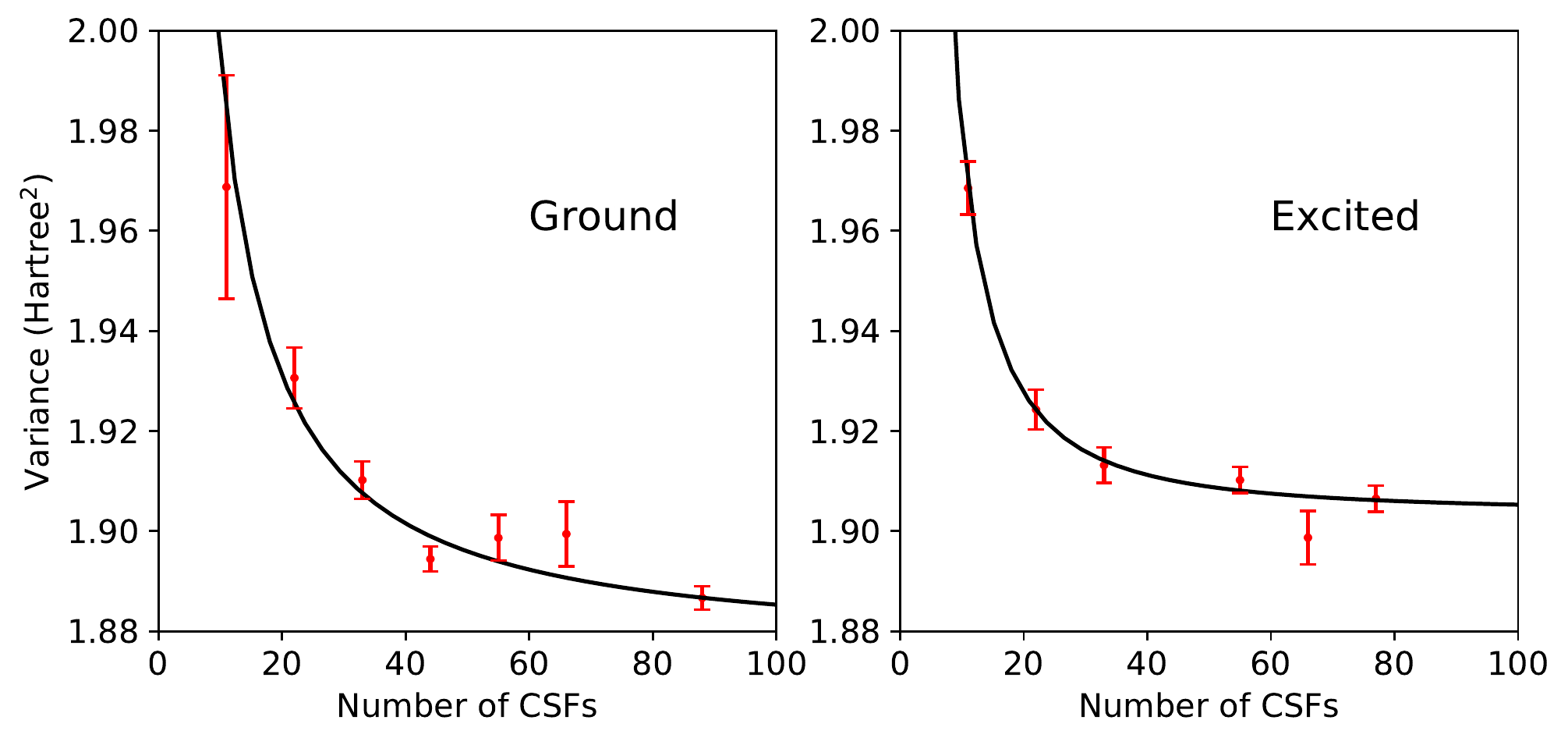}
\caption{Variance estimates for C$_2$ in the ground and first excited singlet state for various MSJ expansion lengths.
         Solid lines are fits based on Eq.\ (\ref{eqn:power_law_fit}).
        }
\label{fig:c2_var}
\end{figure*}

\begin{figure*}
\centering
\includegraphics[width=0.8\textwidth]{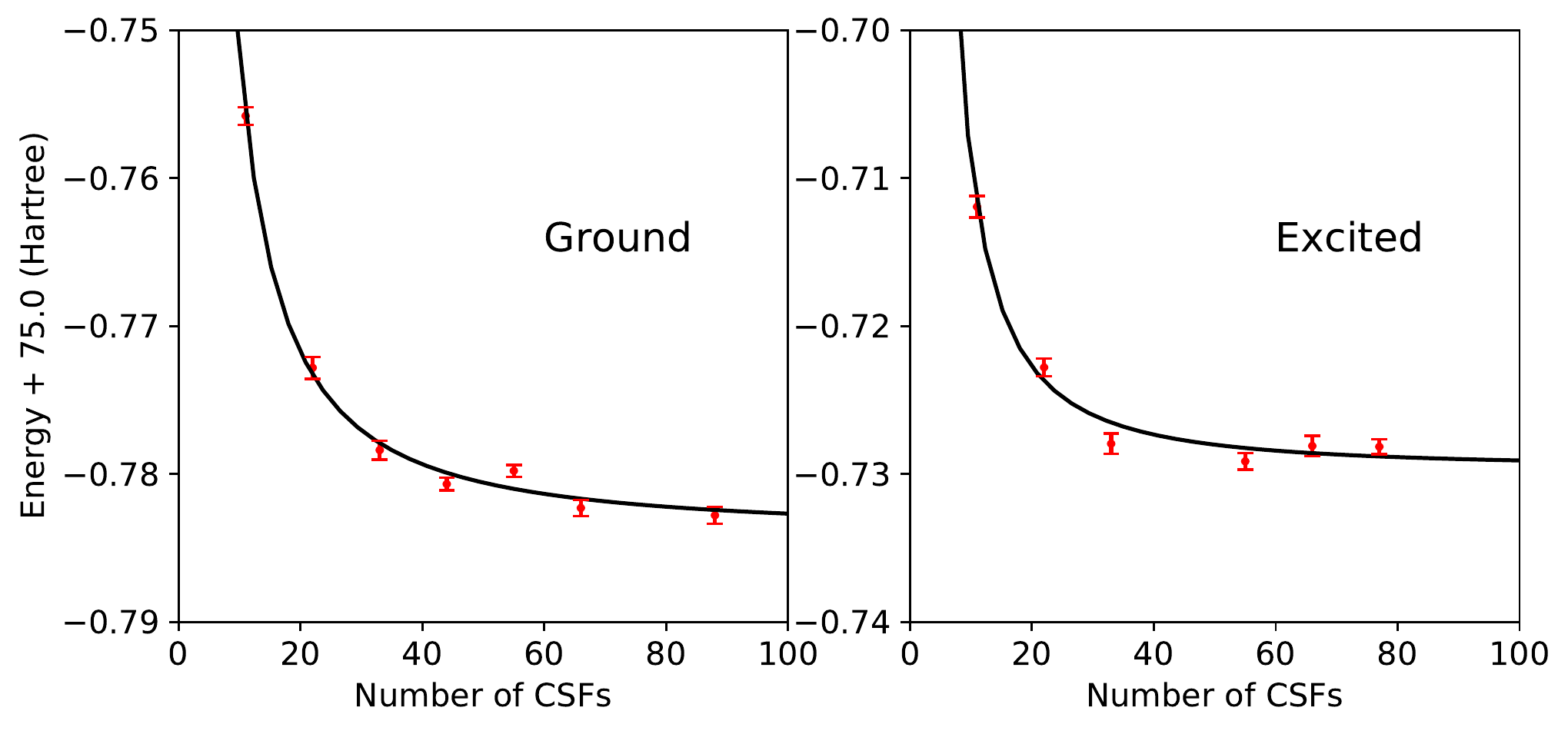}
\caption{Energy estimates for C$_2$ in the ground and first excited singlet state for various MSJ expansion lengths.
         Solid lines are fits based on Eq.\ (\ref{eqn:power_law_fit}).
        }
\label{fig:c2_energy}
\end{figure*}

Specifically, we replace $|\Psi(\vec{r})|^2$ with the guiding function
\begin{align}
p(\vec{r}\hspace{0.5mm}) =
|\Psi(\vec{r}\hspace{0.5mm})|^2
+ \frac{ \epsilon \hspace{0.7mm} |\nabla^2\Psi|^2 }{ 1 + \exp\left( \hspace{0.5mm} \beta \hspace{0.5mm} ( \mathrm{ln}|\Psi| - \gamma ) \right) }
\label{eqn:nodeless_gf}
\end{align}
in order to avoid the divergences in the local energy that cause an infinite variance of the variance.
To keep this distribution as close to $|\Psi(\vec{r})|^2$ as possible when $\vec{r}$ is not
near a node (which is advantageous as it allows us to approach closer to the zero variance principle)
the denominator in Eq.\ (\ref{eqn:nodeless_gf}) works to smoothly ``switch off'' the adjustment
as the logarithm of the wave function rises above the threshold $\gamma$.
While the magnitude of $\nu^2$ will of course depend on the parameters $\epsilon$, $\beta$, and $\gamma$,
it will be finite so long as $\epsilon>0$ and $\beta>0$.
We note that, in practice, evaluating $\nabla^2\Psi$ as part of the sampling function is expensive, and so in future
it will be desirable to test other less expensive alternatives for modified sampling functions.
For the tests in the present study, however, we find this sampling function to be affordable.

\section{Results}
\label{sec:results}

\subsection{Computational Details}
\label{sec::comp_details}

VMC calculations were performed in a development version of QMCPACK \cite{qmcpack_1,qmcpack_2},
with the molecular orbital basis and the choice of CSFs derived from restricted Hartree-Fock (RHF),
complete active space self-consistent field (CASSCF) and CI calculations in GAMESS \cite{gamess1}
as described for each individual molecule in Appendix \ref{apdx:comp_details}.
Equation of motion coupled cluster with singles and doubles (EOM-CCSD),
multi-reference CI with Davidson correction (MRCI+Q),
and complete active space second order perturbation theory (CASPT2)
calculations were carried out with Molpro. \cite{MOLPRO-WIREs}

In all cases, the CSFs used for a given MSJ expansion were taken as the first $N$ CSFs from
a CISDTQ calculation.
While this is of course not the most efficient way to generate a short sCI expansion, it is
sufficient for our purposes of testing the efficacy of variance matching.
The CSF coefficients and Jastrow variables of each MSJ wave function were optimized by minimizing
$\Omega$, with $\omega$ at first held fixed to ensure convergence to the desired state but then adjusted
so as to minimize $\Omega$ for that state (see Reference \cite{Zhao:2016:dir_tar} for details).
The sample length and number of optimization steps were chosen so as to ensure the statistical uncertainty
in the optimized energy was converged to within one or two milliHartrees.
Standard deviations of statistically estimated quantities are displayed in the data below.
In the case of Figure \ref{fig:c2_var} where the central limit theorem does not apply,
it should be understood that although the error bars have been evaluated using the typical
$\sigma/\sqrt{N}$ formula, the underlying statistics cannot be assumed to be Gaussian.

In all cases, the electron-nuclear and electron-electron Jastrow factors were
parameterized as exponentials of 10-point splines with 5-Bohr and 10-Bohr cutoffs, respectively.
Nuclear cusps were enforced by augmenting the orbitals \cite{Needs:2005:cusps_in_orbs},
while electron-electron cusp conditions \cite{Foulkes:2001:qmc_review} were enforced via the electron-electron Jastrow.
For system-specific computational details, such as geometries and basis sets, we refer the reader to Appendix \ref{apdx:comp_details}.

\subsection{C$_2$}
\label{sec::c2}

We begin our investigation of variance matching in the carbon dimer by optimizing
a series of MSJ wave functions with increasingly large CSF expansions for both the
ground and first excited singlet state.
We intentionally restrict ourselves to relatively short expansions with no more
than 100 CSFs so as to simulate a situation in which results cannot be converged
by the brute-force expansion of the wave function, even though this would of course
be possible in a system as small as C$_2$.
As shown in Figures \ref{fig:c2_var} and \ref{fig:c2_energy},
increasing the CSF length has the expected effect of reducing the variance and energy
for both states.

Note that we completed our studies of C$_2$ and allene before implementing the
adjusted importance sampling function of Eq.\ (\ref{eqn:nodeless_gf}), and so these
systems' results are based on variance estimates using the standard $|\Psi|^2$
importance sampling.
As expected, this appears to have had a deleterious effect on the uncertainty in
the variance, but in both cases it was nonetheless possible to perform reasonable
fits for the purposes of variance matching.
We will see that the variance statistics are much better behaved in H$_4$ and C$_3$,
where we have utilized our modified importance sampling.


\begin{figure}[t]
\includegraphics[width=8.0cm,angle=0]{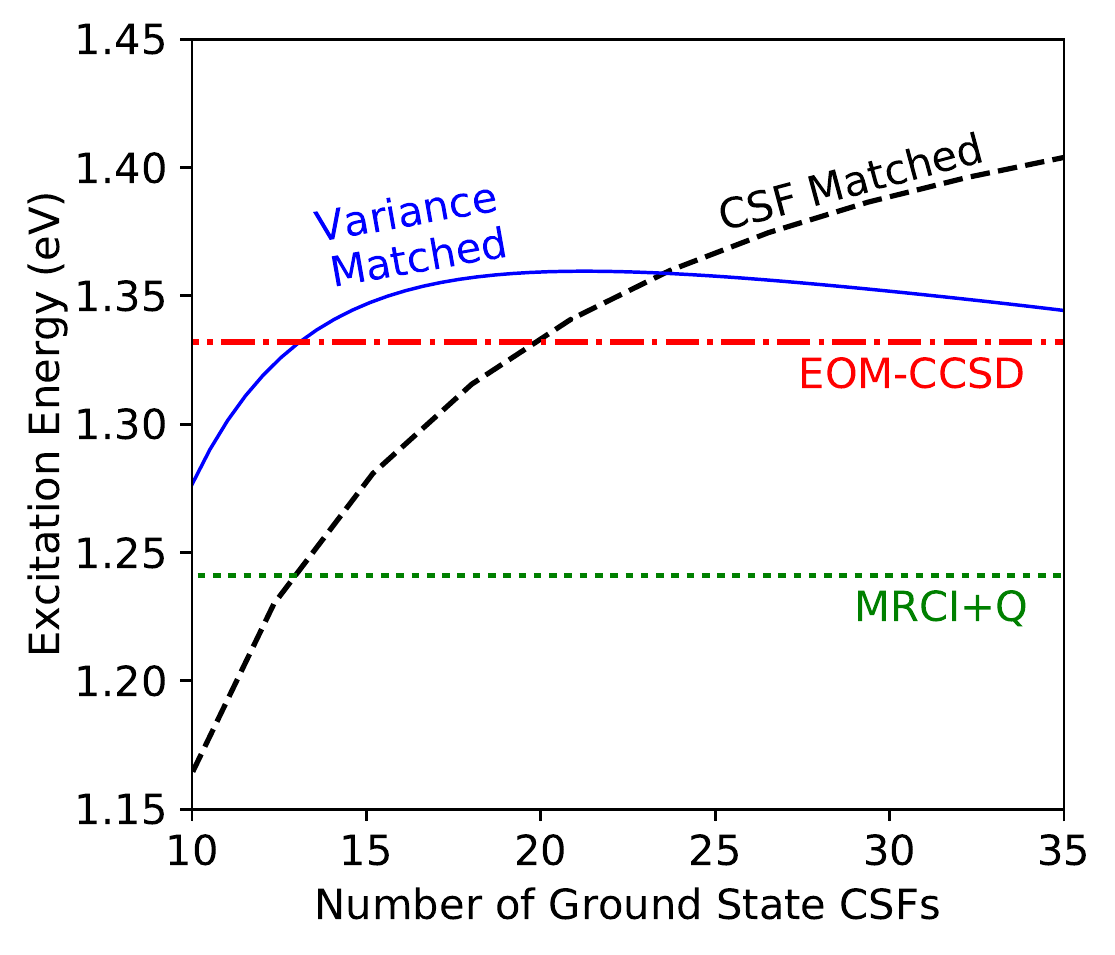}
\caption{Predicted excitation energy for the first excited state of C$_2$ using both
         CSF number matching (in which the excited MSJ had the same number of CSFs as the ground state)
         and variance matching (in which the excited MSJ had the same variance as the ground state).
         Both curves are derived from the interpolations shown in Figures \ref{fig:c2_var} and \ref{fig:c2_energy}.
         Two high-level benchmarks are shown for comparison.
        }
\label{fig:c2_matching}
\end{figure}

\begin{table}[b]
\caption{Excitation energies for allene's first excited singlet.
        }
\centering
\begin{tabular}{lll}
\hline
Method                & Orbitals            & h$\nu$ (eV)        \\
\hline                                     
EOM-CCSD & RHF                 & $6.186$            \\
CASSCF                & CASSCF              & $6.456$            \\
MRCI+Q                & CASSCF \hspace{4mm} & $6.303$            \\
$\sigma^2$-matched ($\sigma^2=3.44$) \hspace{3mm} & CASSCF              & $6.484\pm0.002$    \\
CSF-matched ($N=22$)  & CASSCF              & $7.066\pm0.003$    \\
CSF-matched ($N=66$)  & CASSCF              & $6.651\pm0.003$    \\
\hline
\end{tabular}
\label{tab:alleneData}
\end{table}

Despite the somewhat noisy data,
we can use the fits to Eq.\ (\ref{eqn:power_law_fit}) to
produce an estimate of what the
excitation energy would be
when the excited state MSJ expansion length is chosen so that it has the same variance
as the ground state.
We plot these results in Figure \ref{fig:c2_matching} alongside the excitation energies
that result when both states are restricted to have the same number of CSFs, which,
despite enforcing equality in flexibility, is ineffective at balancing accuracy
as the expansion length needed to handle the effects of correlation in different states
is of course not uniform.
Indeed,
we see that the size of the ground state CSF expansion has a smaller effect
on the predicted excitation energy under variance matching, with predictions all falling
within 0.09eV of each other for ground state CSF lengths between 10 and 35 CSFs.
If instead we match the number of CSFs in each state, the predicted excitation energy
changes by 0.24eV over this range.
Furthermore, the variance-matched results appear to converge more rapidly
towards agreement with the benchmark EOM-CCSD and MRCI+Q results 
as the number of ground state CSFs is increased.
Thus, as hoped, the use of variance matching for this excitation appears to help offset
the fact that the two states require significantly different numbers of CSFs to
reach descriptions of equivalent accuracy.

\begin{figure*}
\centering
\includegraphics[width=0.8\textwidth]{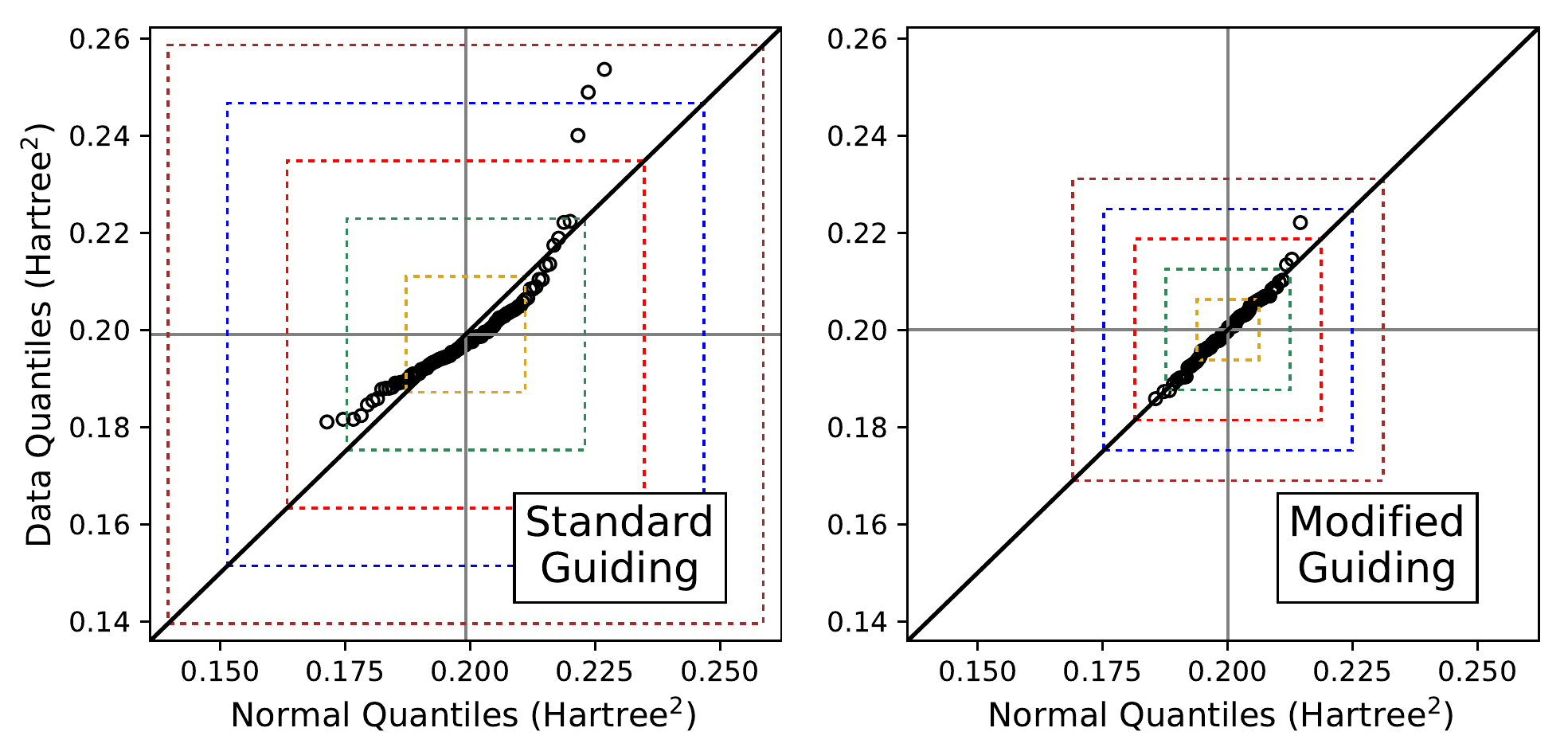}
\caption{Quantile-quantile plots comparing the observed distribution of H$_4$ variance estimates (data quantiles)
         to the average distribution that would be expected for the variance if it were normally distributed
         with the same mean and standard deviation (normal quantiles).
         Each plot contains 100 circles, each of which marks the variance as estimated over an independent random sample
         of 16,384 electron configurations drawn from either the standard guiding function $|\Psi(\vec{r})|^2$
         or the modified function of Eq.\ (\ref{eqn:nodeless_gf}).
         Dashed squares show the bounding regions for the first five standard deviations of the observed distributions.
         Solid vertical and horizontal lines show the mean, while the slanted black line
         shows where data points would be expected to lie if they exactly matched the quantiles of a normal distribution.
        }
\label{fig:h4_qq}
\end{figure*}

\subsection{Allene}
\label{sec::allene}

The second system we investigate is allene, C$_3$H$_4$.
Unlike C$_2$, in which the ground state itself is quite multi-reference,
allene presents us with a case in which the ground and first $\pi \to \pi^*$ excitation have significantly different
degrees of multi-reference character.
The ground state of allene is dominated by the closed shell determinant: its CI coefficient is 0.96 in CASSCF
and only 3 CASSCF CI coefficients are greater than 0.001.
In contrast, the first excited state has 14 CASSCF CI coefficients above this threshold, making allene an excellent candidate for
a system in which variance matching should be helpful.

As with C$_2$, we limited ourselves to very short CSF expansions and found that when
the ground state was optimized with a 22-CSF MSJ, the variance was 3.44 Hartree$^2$.
After some initial trial-and-error and interpolation, we found that the same variance
results for the excited state when a 66-CSF MSJ expansion is employed.
In Table \ref{tab:alleneData},
we see that when MSJ expansions with the same number of CSFs are used for both the
ground and excited state, the excitation energy is overestimated as compared to the
EOM-CCSD and MRCI+Q results, which is what we would expect given the more
multi-reference character of the excited state.
CSF-matched results do improve as the number of CSFs is increased from 22 to 66,
but we see that it is more accurate to evaluate the excitation energy based on
a ground state with 22 CSFs and an excited state with 66, which as stated above
produces states with equal values of $\sigma^2=3.44$ Hartree$^2$.
Thus, as expected, variance matching for this excitation reduces the ground state bias that originates
from having a more multi-reference excited state.

\subsection{H$_4$}
\label{sec::h4}

At this point in our investigation, we decided that although variance matching
looked promising, the poor statistics of the variance should be addressed
before proceeding further.
To this end, we implemented importance sampled Monte Carlo integration based
on Eq.\ (\ref{eqn:nodeless_gf}) in our development version of QMCPACK.
To validate that the infinite variance of the variance issue was
satisfactorily resolved and the statistics of the variance were now normal
as per the central limit theorem, we performed normality tests for
variance data on a simple system of two nearby hydrogen molecules,
one with a near equilibrium bond
length and the other slightly stretched, a system we will refer to as H$_4$.
After optimizing the MSJ ansatz for the ground state,
we interrogated the statistical distribution of its variance by
estimating it with multiple, independent samples of electron
configurations, the results of which are displayed in
Figure \ref{fig:h4_qq}.

\begin{figure*}
\centering
\includegraphics[width=0.8\textwidth]{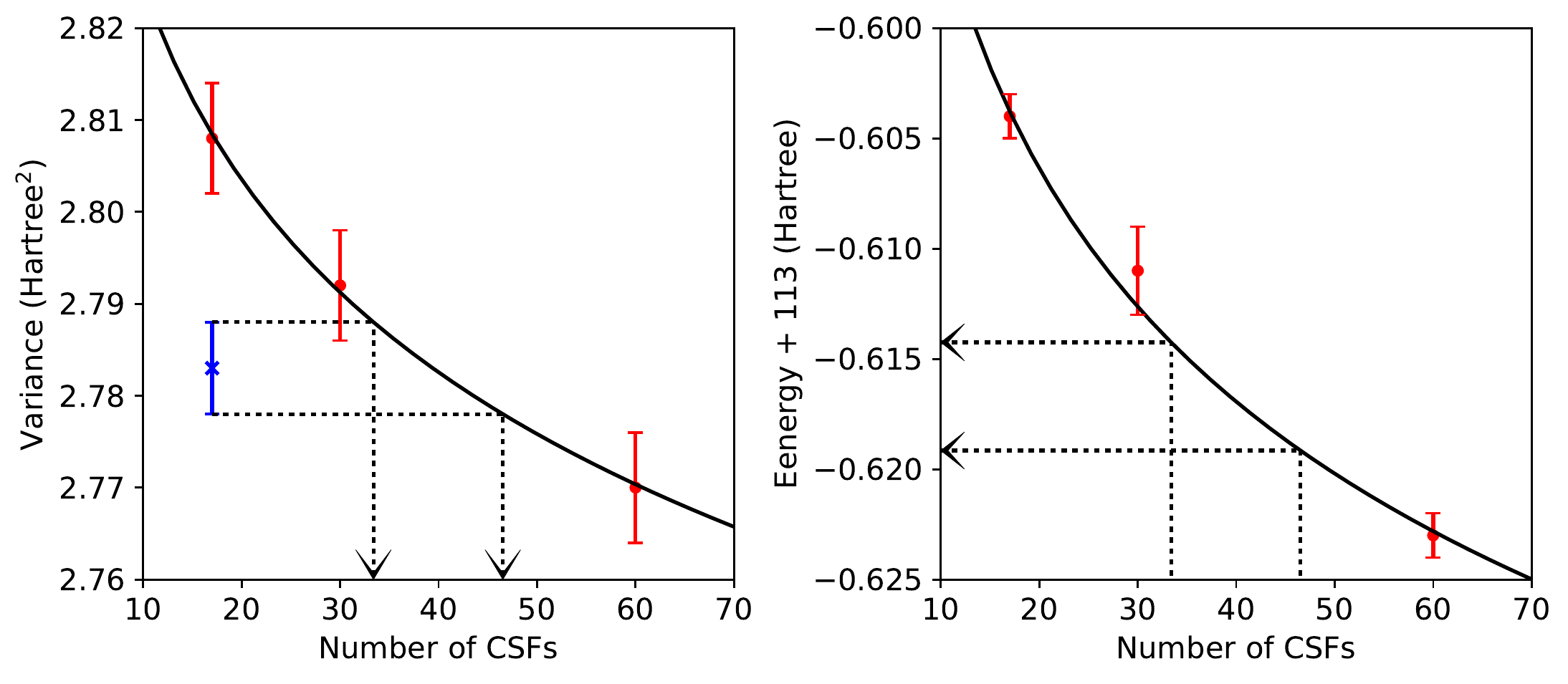}
\caption{Variance and energy estimates for the ground (blue x) and first excited singlet state (red circles)
         of C$_3$ when working in the RHF molecular orbital basis.
         Solid lines show fits of the excited state data to Eq.\ (\ref{eqn:power_law_fit}), while dashed lines
         show the upper and lower bounds of the excited state expansion lengths and corresponding energies
         that give variances within the error bars of the ground state variance.
         See Section \ref{sec::c3} for details.
        }
\label{fig:c3_rhf}
\end{figure*}

\begin{figure*}
\centering
\includegraphics[width=0.8\textwidth]{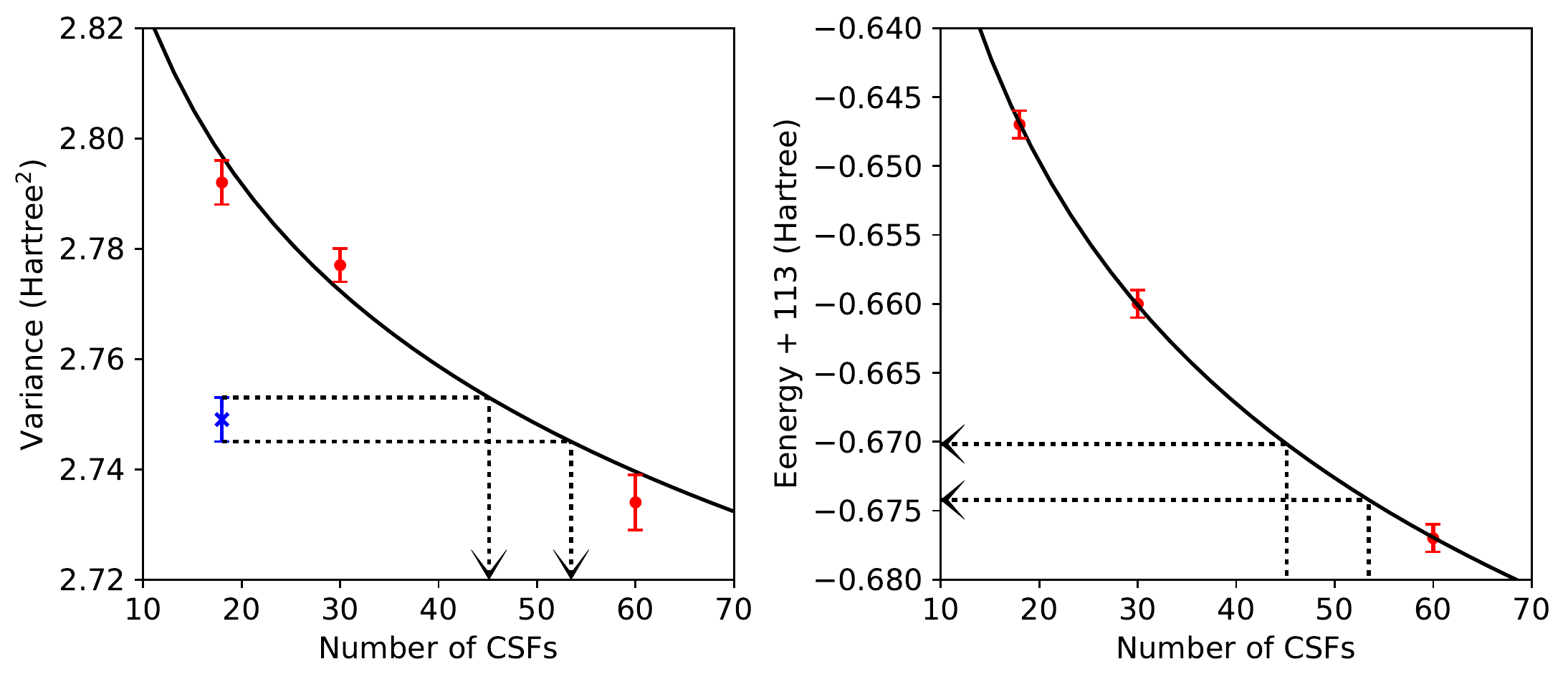}
\caption{As for Figure \ref{fig:c3_rhf}, but now using CASSCF molecular orbitals.
        }
\label{fig:c3_casscf}
\end{figure*}

\begin{figure*}
\centering
\includegraphics[width=0.8\textwidth]{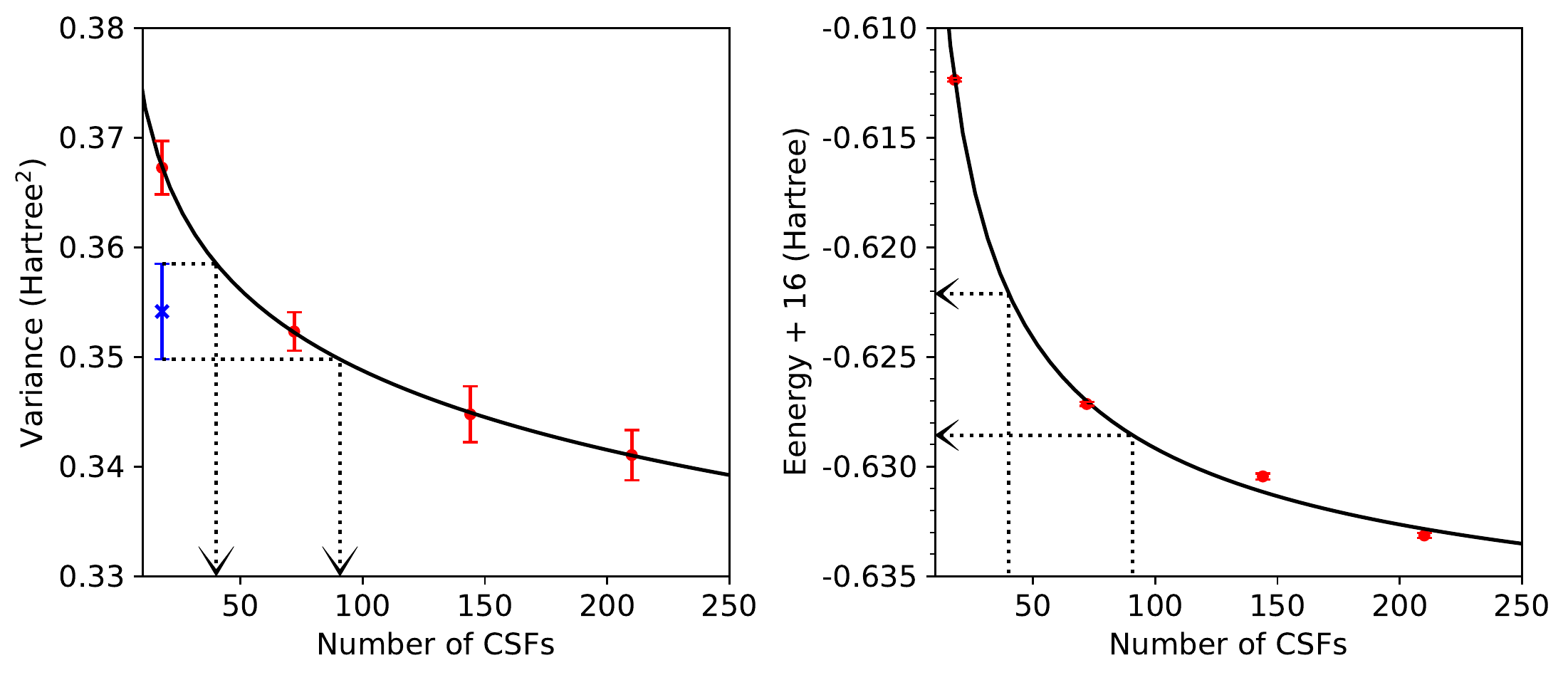}
\caption{As for Figure \ref{fig:c3_rhf}, but now using an effective core potential and VMC-optimized orbitals.
        }
\label{fig:c3_opt}
\end{figure*}

The statistical distribution of $\sigma^2$ is clearly skewed away
from normal when using the standard $|\Psi|^2$ guiding function,
confirming that the central limit theorem is not in effect,
as expected. \cite{Trail:2008:heavy_tail}
This conclusion is further buttressed by the observation that
two out of the 100 estimates taken with $|\Psi|^2$ lie more than
four standard deviations away from the mean.
On average, normally distributed data would only show two
4-$\sigma$ events like this after 31,574 estimates, and here we
see two after just 100.

Using the modified guiding function, the variance estimates match
a normal distribution much more closely, and we observe no 4-$\sigma$ events.
We do see one 3-$\sigma$ event among the 100 estimates, compared
to an expectation that, on average, one 3-$\sigma$ event will occur in every
370 estimates for normally distributed data.
This deviation is much more believably either a statistical fluctuation
or the effect of the fact that our Monte Carlo samples of configurations are
finite and so we may not yet have converged to the perfectly Gaussian limit.
In addition to becoming more normal in their distribution, the statistical
uncertainties in the variance are also smaller with the modified guiding function,
at least if we measure by their standard deviation.
Thus, as previously predicted, \cite{Trail:2008:heavy_tail,Trail:2008:alt_sampling}
it appears that we can indeed resolve the infinite variance of the variance
dilemma via a modification to the importance sampling function that we 
employ in our Monte Carlo integration.

\subsection{C$_3$}
\label{sec::c3}

\begin{table}[b]
\centering
\caption{Excitation energies for C$_3$'s first excited singlet.}
\begin{tabular}{lll}
\hline
Method                               \hspace{3mm} & Orbitals \hspace{3mm} & h$\nu$ (eV) \\
\hline
EOM-CCSD                             \hspace{3mm} & RHF      \hspace{3mm} & 3.402 \\
CASSCF                               \hspace{3mm} & CASSCF   \hspace{3mm} & 3.303	\\
CASPT2                               \hspace{3mm} & CASSCF   \hspace{3mm} & 2.871	\\
MRCI+Q                               \hspace{3mm} & CASSCF   \hspace{3mm} & 3.168 \\
CSF-matched                          \hspace{3mm} & RHF      \hspace{3mm} & 4.46 $\pm$ 0.05 \\
CSF-matched                          \hspace{3mm} & CASSCF   \hspace{3mm} & 3.90 $\pm$ 0.04 \\
$\sigma^2$-matched                   \hspace{3mm} & RHF      \hspace{3mm} & 4.12 $\pm$ 0.07 \\
$\sigma^2$-matched                   \hspace{3mm} & CASSCF   \hspace{3mm} & 3.21 $\pm$ 0.06 \\
$\sigma^2$-matched                   \hspace{3mm} & VMC      \hspace{3mm} & 3.14 $\pm$ 0.09 \\
\hline
\end{tabular}
\label{tab:C3_data}
\end{table}

Our final test system in this preliminary investigation of variance matching is
C$_3$, the linear carbon trimer, in which we use the improved variance
statistics granted by the modified guiding function to help resolve
the importance of the choice of molecular orbital basis.
In a setting where a sCI expansion can be converged with respect to CSF number,
the molecular orbital basis will of course not matter.
However, in the more practical case we are considering here, in which converging the
expansion is prohibitively expensive, this guarantee cannot be relied upon.
We address this issue by testing variance-matching
in three sets of molecular orbitals: RHF-optimized orbitals,
CASSCF-optimized orbitals, and VMC-optimized orbitals.
Note that for the latter case, we have optimized the orbitals separately for
the ground state and the first excited state (see Appendix \ref{apdx:comp_details})
using a recently-implemented combination 
of our direct targeting method and the efficient MSJ orbital
optimization approach of Filippi and coworkers. \cite{Filippi:2016:msj_gradients,assaraf2017optimizing}

For each choice of basis, we first performed a ground state MSJ optimization
with a short CSF expansion.
Next, we performed excited state optimizations with different CSF expansion
lengths so that we could estimate by interpolation the excited state expansion
length (and its corresponding energy) at which the ground and excited state
variances would match.
This process and its results are displayed graphically in Figures \ref{fig:c3_rhf}-\ref{fig:c3_opt}
for our three different choices for the molecular orbitals.
Before analyzing the effects of variance matching, it is worth noting that,
thanks to the improved guiding function, the variance uncertainties in this case
are smaller than they were in C$_2$, despite the 50\% larger system size.

As anticipated, Table \ref{tab:C3_data} reveals that the choice of the molecular orbitals
matters when taking a variance-matching approach to excitation energy prediction.
We see that results using RHF orbitals are quite poor, which is to be expected due to the
bias they create in favor of the ground state.
On the other hand, excitation energies based on CASSCF
or VMC-optimized orbitals are within statistical uncertainty of the benchmark
(cc-pVTZ, full-valence-CAS) MRCI+Q results.
Although one cannot expect to have access to CASSCF orbitals in large systems,
it is possible \cite{assaraf2017optimizing} to produce VMC-optimized orbitals, making their
excellent performance in conjunction with variance matching and short CSF expansions
an extremely promising result.
As a final note, and as we saw in other cases, CSF-number-matched results are less accurate,
reinforcing the point that providing equal amounts of wave function flexibility
is no guarantee that the errors in different states will be similar.
Instead, it proves far more effective to ensure that the states are balanced in terms
of a rigorous error metric, such as $\sigma^2$.

\section{Conclusions}
\label{sec:conclusion}

Extending the success of sCI methods into regimes where exhaustive convergence is not possible
would greatly expand their utility in chemical investigations.
Towards this end, we have developed and tested a strategy for excitation energies based
on variational Monte Carlo's ability to use the energy variance to balance the descriptions
of different states.
In tests on C$_2$, allene, and C$_3$, our variance-matched results are within 0.1, 0.2, and 0.1 eV,
respectively, of cc-pVTZ MRCI+Q excitation energies.
These accuracies are achieved despite using ground state selective configuration
interaction expansions that
contain fewer than 40 configuration state functions and are thus
very far from the exhaustive flexibility limit.
As multi-Slater Jastrow optimizations can now handle hundreds of thousands of determinants and
molecules containing over 100 atoms,
\cite{Filippi:2016:msj_gradients,assaraf2017optimizing}
these preliminary results suggest that variance matching could play a major role in future
high-accuracy work in molecular excited states.

To make further progress towards this goal, it will be important to focus improvement on the most
important aspects of the methodology.
As we showed in this study, statistical estimates of the variance are greatly improved by using
a non-standard guiding function for the Monte Carlo integration that, although it has the
same scaling with system size, is more expensive to work with than the standard $|\Psi|^2$ guiding function.
It would thus be beneficial to pursue more sophisticated guiding functions that are both
inexpensive and effective at reducing uncertainty in variance estimates.
Another avenue of investigation that will clearly be important is the choice of selective CI
scheme.
While many options exist, none are currently native to quantum Monte Carlo, by which we mean that they do not
take account of the Jastrow correlation factor when making decisions about
which configurations to use.
Developing selection schemes that couple more closely to quantum Monte Carlo is therefore highly desirable.
By pursuing improvements in these areas, and by pushing to larger systems and more aggressive determinant
expansions, we look forward to future explorations of how variance matching can complement existing methods
in achieving balanced and accurate treatments of molecular excited states.

\section{Acknowledgements}

The authors acknowledge funding from the Office of Science, Office of Basic Energy Sciences,
the US Department of Energy, Contract No.\ DE-AC02-05CH11231.
Parallel development and debugging of the software for nodeless guiding functions was performed
using the Berkeley Research Computing Savio cluster.
Molecular calculations were run at the National Energy Research Scientific Computing Center,
a DOE Office of Science User Facility supported by the Office of Science of the U.S. Department
of Energy under Contract No.\ DE-AC02-05CH11231.

%

\appendix

\section{Modified Guiding Example}
\label{apdx:mge}

To see how the breakdown of the central limit theorem comes about and how this
difficulty can be addressed by a modified guiding function, consider a two-dimensional integral
\begin{align}
\label{eqn:example_func}
f(x,y) = \frac{\mathrm{cosh}\left(4\sqrt{x^2+y^2}\right) e^{-x^2-y^2}}{1000}
\end{align}
\begin{align}
\label{eqn:example_int}
\mu = \int_{-\infty}^{\infty} \int_{-\infty}^{\infty} f(x,y) \hspace{0.5mm} dx \hspace{0.5mm} dy \hspace{0.5mm} \approx \hspace{0.5mm} 0.608
\end{align}
which we might try to evaluate via importance-sampled Monte Carlo integration.
For a given choice of normalized importance-sampling function $p(x,y)$, we have
\begin{align}
\label{eqn:mci_g}
& g(x,y)=\frac{f(x,y)}{p(x,y)} \\
\label{eqn:mci_int}
& \mu = \big\langle g \big\rangle_p \equiv \int_{-\infty}^{\infty} \int_{-\infty}^{\infty} p(x,y) \hspace{0.5mm} g(x,y) \hspace{0.5mm} dx \hspace{0.5mm} dy.
\end{align}
The variance for this choice of sampling function is
\begin{align}
\label{eqn:mci_v}
\sigma^2 = \big\langle ( g - \mu )^2 \big\rangle_p
\end{align}
and the variance of the variance is
\begin{align}
\label{eqn:mci_v_of_v}
\nu^2 = \big\langle \big( ( g - \mu )^2 - \sigma^2 \big)^2 \big\rangle_p.
\end{align}

Given that the function $f(x,y)$ decays rapidly, an excellent approximation to the integral
can be had by choosing $p(x,y)$ to be uniform on the square $-10<x<10$, $-10<y<10$.
This produces the mean, variance, and variance of the variance
\begin{align}
\mu = 0.608 \qquad \sigma^2 = 4.325 \qquad \nu^2 = 285.299.
\label{eqn:m_v_vv_unif}
\end{align}
If instead we choose our sampling function as
\begin{align}
p(x,y) = c \hspace{0.5mm} \sqrt{x^2+y^2} \hspace{0.8mm} e^{-(x^2+y^2)/6} ,
\label{eqn:div_imp_func}
\end{align}
with $c$ the normalization constant, we have
\begin{align}
\mu = 0.608 \qquad \sigma^2 = 0.149 \qquad \nu^2 = \infty.
\label{eqn:m_v_vv_div}
\end{align}
Although this importance sampling function reduces the variance, which will
in turn reduce the statistical uncertainty in $\mu$, it creates a singularity
in $g(x,y)$ at the origin.
For $\sigma^2$, this singularity is integrable.
For $\nu^2$, which contains a higher power of $g$, it is not.
We are therefore unable to rely on the central limit theorem in this case
to analyze the uncertainty in a Monte Carlo integration estimate of the variance,
as the variance of the variance is not finite.

This issue is readily resolved via a small modification to the sampling function,
\begin{align}
p(x,y) = c \hspace{0.5mm} \left(\frac{1}{20}+\sqrt{x^2+y^2}\right) \hspace{0.8mm} e^{-(x^2+y^2)/6},
\label{eqn:reg_imp_func}
\end{align}
which removes the singularity in $g$ while only slightly altering its values away from the origin.
The result,
\begin{align}
\mu = 0.608 \qquad \sigma^2 = 0.148 \qquad \nu^2 = 0.013,
\label{eqn:m_v_vv_div}
\end{align}
is that the variance is essentially unchanged, but the variance of the variance is now finite
as there are no longer any singularities.
Thus, in this example, we see that a small change to the sampling function
that removes singularities while otherwise leaving the function more or less unaltered
leads to a qualitative improvement in statistical estimation.
For a rigorous analysis of these issues in VMC, we refer the reader to the work
of Trail.  \cite{Trail:2008:heavy_tail,Trail:2008:alt_sampling}

\section{Computational Details}
\label{apdx:comp_details}

\noindent
\textit{C$_2$}:\hspace{2mm}
The C$_2$ bond length was chosen to be 1.2425146399 \AA.
The basis set was chosen as cc-pVTZ for both the CASSCF and the high-level reference calculations. 
The molecular orbitals were obtained from an (8e,8o) state-averaged CASSCF for the ten lowest lying singlet states.
The configuration list from which truncations were taken for MSJ expansions was generated via a CISDTQ calculation in which
the 1s orbitals were held frozen and excitations were allowed among the 3rd through 22nd CASSCF orbitals.
\vspace{4mm}

\noindent
\textit{Allene}:\hspace{2mm}
The geometry for allene (C$_3$H$_4$),
\begin{tabular}{l r @{.} l r @{.} l r @{.} l}
\hspace{2mm} C  & \hspace{1.6mm} 0&000000000  & \hspace{1.6mm} -0&000000000 &  \hspace{1.6mm}  -0&000000012 \\
\hspace{2mm} C  & \hspace{1.6mm} 0&000000000  & \hspace{1.6mm} -0&000000000 &  \hspace{1.6mm}   2&481842730 \\
\hspace{2mm} C  & \hspace{1.6mm} 0&000000000  & \hspace{1.6mm}  0&000000000 &  \hspace{1.6mm}  -2&481842716 \\ 
\hspace{2mm} H  & \hspace{1.6mm} 1&756493809  & \hspace{1.6mm}  0&000000000 &  \hspace{1.6mm}  -3&532793008 \\
\hspace{2mm} H  & \hspace{1.6mm}-1&756493809  & \hspace{1.6mm}  0&000000000 &  \hspace{1.6mm}  -3&532793008 \\
\hspace{2mm} H  & \hspace{1.6mm} 0&000000000  & \hspace{1.6mm}  1&756493826 &  \hspace{1.6mm}   3&532792993 \\
\hspace{2mm} H  & \hspace{1.6mm} 0&000000000  & \hspace{1.6mm} -1&756493826 &  \hspace{1.6mm}   3&532792993
\end{tabular}
was chosen based on a CCSD(T) optimization with the cc-pVTZ basis set in Molpro. 
The CASSCF orbitals were obtained from a 4-state-averaged CASSCF(10e,16o) calculation in GAMESS with the cc-pVTZ basis.
The CSFs used for the MSJ expansion were generated from a 4-state CISDTQ calculation 
with the 1s orbitals on carbon frozen and excitations allowed among the 4th through 20th CASSCF orbitals.
\vspace{4mm}

\noindent
\textit{H$_4$}:\hspace{2mm}
This system consists of two nearby hydrogen molecules, one with its bond somewhat
stretched so as to induce a degree of correlation in the ground state, for which
we optimized a 2-CSF MSJ wave function.
The basis set was STO-3G, with the molecular orbital basis
for the CI expansion taken as the RHF canonical orbitals.
The atomic coordinates are:
\begin{tabular}{r @{.} l r @{.} l r @{.} l}
\hspace{2mm} 0&0000000000  & \hspace{1.6mm} 0&0000000000  & \hspace{1.6mm} 0&0000000000 \\
\hspace{2mm} 1&8897259877  & \hspace{1.6mm} 0&0000000000  & \hspace{1.6mm} 0&0000000000 \\
\hspace{2mm} 0&0000000000  & \hspace{1.6mm} 0&0000000000  & \hspace{1.6mm} 2&8345889816 \\
\hspace{2mm} 0&0000000000  & \hspace{1.6mm} 0&0000000000  & \hspace{1.6mm} 5&6691779632
\end{tabular}
\vspace{4mm}

\noindent
\textit{C$_3$}:\hspace{2mm} 
The carbon atoms were arranged in a line with 1.302 \AA\ between atoms.
All QMC calculations used the cc-pCVTZ basis set \cite{Dunning:1989:basis_for_corr}
to help reduce variance in the core region,
while reference calculations were performed in the more typical cc-pVTZ basis.
The reference CASSCF, CASPT2, and MRCI+Q calculations used a (12e,12o) ``full valence'' active space.
In EOM-CCSD, CASPT2, and MRCI+Q, the 1s orbitals were set as an inactive frozen core.

To generate our multi-Slater expansion, 
we employed molecular orbitals from either RHF or an equally-weighted CASSCF state average
of the first three singlet states in an (8e,10o) active space.
For a given choice of CSF number $N$, the MSJ's CI expansion was constructed using
the $N$ CSFs with the largest coefficients after a single-reference CISDTQ calculation
(in either the RHF or CASSCF orbitals)
in which the 1s orbitals were held frozen and excitations were allowed among the 4th through
27th molecular orbitals.
Note that for the ground state, we used 17 CSFs in the RHF orbital case and 18 CSFs
in the CASSCF orbital case, as these choices avoided splitting degenerate CSFs
between the MSJ's used and unused CSF sets.

The VMC optimized molecular orbitals were found by simultaneously optimizing the orbitals,
Jastrow variables, and CI coefficients for 18-CSF MSJ wave functions for the ground
and excited states separately.
To aid convergence in these optimizations, we have replaced the 1s electrons
with an effective core potential \cite{carbon_pseudopotential} in all calculations involving VMC-optimized orbitals.
For consistency, we took our CSFs in this case from the closest possible CISDTQ,
which was generated in the same way as for the CASSCF orbital case, except that the 
effective core potential was employed.

\bibliographystyle{aip}

\end{document}